\documentclass[a4paper,11pt]{article}
\pdfoutput=1 

\usepackage{jinstpub} 
\usepackage{mwe}
\usepackage{subfig}

\title{\boldmath Thin and edgeless sensors for ATLAS pixel detector upgrade}

\author[a,1]{Audrey Ducourthial,\note{Corresponding author.}}
\author[a]{Marco Bomben,}\author[a]{Giovanni Calderini,}\author[a]{Giovanni Marchiori,}\author[a]{ Louis D'Eramo,}\author[a]{Ilaria Luise,}\author[b]{Alvise Bagolini,}\author[b,c]{ Maurizio Boscardin,}\author[d]{Luciano Bosisio,}\author[g]{Giovanni Darbo,}\author[c,f]{Gian-Franco Dalla Betta,}\author[b,e]{Gabriele Giacomini}\author[h]{Marco Meschini,}\author[i]{Alberto Messineo,}\author[b,c]{Sabina Ronchin,}\author[b,c]{Nicola Zorzi}

\affiliation[a]{Laboratoire de Physique Nucleaire et de Hautes Energies, LPNHE,4 place Jussieu, Paris, France}
\affiliation[b]{Fondazione Bruno Kessler, Centro per i Materiali e i Microsistemi (FBK-CMM), I-38123 Povo di Trento (TN), Italy}
\affiliation[c]{Trento Institute for Fundamental Physics and Applications (TIFPA INFN),Trento, Italy}
\affiliation[d]{Universita di Trieste, Dipartimento di Fisica and INFN, I-34127 Trieste, Italy}
\affiliation[e]{Brookhaven National Laboratory, Instrumentation Division 535B, Upton, NY - USA}
\affiliation[f]{University of Trento, INFN}
\affiliation[g]{INFN Genova}
\affiliation[h]{ INFN Firenze}
\affiliation[i]{ Universita Pisa (IT), INFN Pisa}
\emailAdd{audrey.ducourthial@cern.ch}

\abstract{To cope with the harsh environment foreseen at the high luminosity conditions of HL-LHC, the ATLAS pixel detector has to be upgraded to be fully efficient with a good granularity, a maximized geometrical acceptance and an high read out rate. LPNHE, FBK and INFN are involved in the development of thin and edgeless planar pixel sensors in which the insensitive area at the border of the sensor is minimized thanks to the active edge technology. In this paper we report on two productions, a first one consisting of 200 $\mu m$ thick n-on-p sensors with active edge, a second one composed of 100 and 130 $\mu m$ thick n-on-p sensors. Those sensors have been tested on beam, both at CERN-SPS and at DESY and their performance before and after irradiation will be presented.}


\arxivnumber{1234.56789} 

\proceeding{11$^{\text{th}}$ Conference on Position Sensitive Detectors\\
  Milton Keynes\\
  3-8 september 2017}

\begin{document}
\maketitle
\flushbottom

\section{ATLAS tracker upgrade}
\label{sec:intro}
\subsection{ATLAS inner tracker}
\begin{figure}[htbp]
\centering
\includegraphics[scale=0.8]{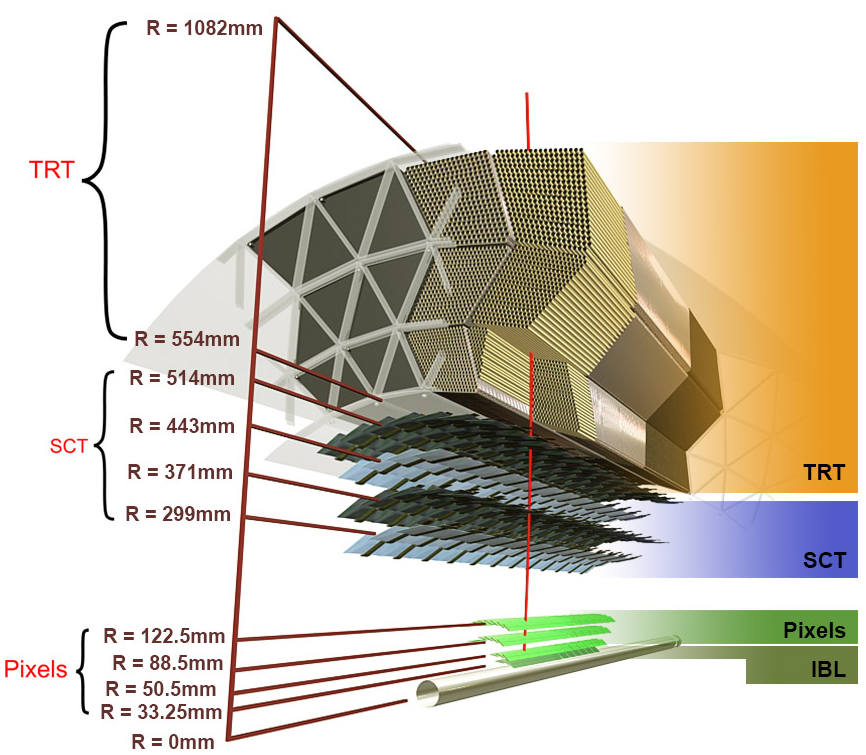}
\caption{\label{fig:1} Inner ATLAS detector with Run2 layout}
\end{figure}

The ATLAS~\cite{atlas} inner detector  ~\cite{atlasid1} is dedicated to the study of tracks from particles created at the point of interaction between two partons.
It is composed of three sub-detectors: a transition radiation tracker, a strip detector and a pixel detector~\cite{atlasid2}, the closer to the beam pipe which is subdivided into a barrel part in the center and two endcap parts at each end of the detector. The barrel pixel detector is now running with 4 pixel layers, whose distances to the beam are documented in figure \ref{fig:1}.
The first layer, namely the IBL (Insertable B Layer) ~\cite{ibl} is of special help to distinguish tracks from B hadrons decays, which is crucial for many physics analyses. Its proximity to the interaction point makes this detector particularly concerned by radiation damage.
The three outer layers and $75\%$ of the IBL are made of planar pixel sensors, which have demonstrated excellent performance during the Run 1 and Run 2, taking data in a harsh radiative environment (the current fluence is of the order of $ 10^{14}n_{eq}/cm^2$ and the expected end-of-lifetime fluence of the IBL is $5\times 10^{15}n_{eq}/cm^2$) at a 20 MHz rate with a spatial resolution of about 10 micrometers.

\subsection{ATLAS inner tracker upgrade}
During the data taking period of the High Luminosity LHC (HL-LHC)\cite{hllhc}, the ATLAS detector will have to take data in extreme conditions and is expecting:
\begin{itemize}
		\item an instantaneous peak luminosity  of the order of  L$\simeq 7.5\times10^{34}cm^{-2}s^{-1}$, a five-fold increase with respect of today;
		\item 200 inelastic proton proton  collisions per bunch crossing;
		\item an accumulated luminosity of 4000 $fb^{-1}$ by the end of 2035, which means that the innermost tracker will be exposed to a fluence of $2\times 10^{16} n_{eq}/cm^2$ (4 times the IBL fluence).
	\end{itemize}
To cope with the high luminosity conditions of the HL-LHC, the ATLAS inner tracker upgrade will have to face three major challenges:
 \begin{itemize}
 \item Radiation hardness: retain a 97$\%$ efficiency with an expected fluence up to $1\times 10^{16} neq/cm^2$ for the inner most layer. 3D pixel sensors~\cite{ibl} are the preferred option for the innermost layer, as they are intrinsically more radiation hard than planar pixel sensors with a shorter collection distance. For the outermost layers, thin planar pixel sensors ($\simeq 100\mu m$ thick) are developed because they are less sensitive to charge trapping than the traditional thicker planar pixel sensors ($\simeq 200\mu m$ thick).
 
 \item Pile up compliance: due to the high luminosity conditions, more particles are expected to cross a pixel sensor. To ensure a good reconstruction of tracks in this dense environment, the spatial resolution must be increased, hence the granularity of the pixel cell will be increased. The actual pitch of the sensor is $250\mu m \times 50 \mu m$, it will be reduced to $ 50 \mu m \times 50 \mu m$ or $25 \mu m \times 100 \mu m$. A new chip, RD53~\cite{rd53}, with a pitch of $50\times 50 \mu m$ is also in development.
 \item Increase the geometrical acceptance: to maximize the active part of the detector, the new detector will be instrumented at large pseudorapidity $\eta$\footnote{ $\eta \equiv -\ln \left[\tan \left({\frac {\theta }{2}}\right)\right]$ is the angular dependency notation of the ATLAS detector }, up to $ \lvert \eta \lvert =4$. To ensure good inter pixel module coverage, the dead area at the edge of each module has to be kept at minimum~\cite{strips}. This can be achieved by using the active edge technology (see section \ref{sec:atcive}).

\end{itemize}

Thin and edgeless planar pixel sensors are then good candidates to fulfill the requirement of the upgraded ATLAS pixel detector.

\section{LPNHE FBK INFN sensors}

The LPNHE in collaboration with the FBK foundry and INFN has developed a sequence of 3 planar pixel sensor productions. The first two productions have been tested on beam. The first production is composed of active edge pixel sensors, the second of thin sensors and the third one combines the technology of the first two.

\subsection{First production: active edge sensors  }
\label{sec:atcive}
The first production consists of n-on-p devices with a bulk thickness of $200\mu m$ \cite{bomben}. The pitch of those pixels is $250 \mu m \times 50 \mu m$, compatible with the current IBL read out chip, the FEI4b~\cite{fei4}.
The active edge technology~\cite{active} consists in etching a passing-through trench via a Deep Reactive Ion Etching (DRIE) process at the border of the sensor and in doping the walls of the excavated trenches. It allows to reduce the extension of the dead area, by constraining the border of the depletion region at the  trench edge and by keeping out the cracks that can be caused by a traditional diamond saw cut. To fully exploit this technique, the number of Guard rings must be considerably reduced. The tested sensors had 0 to 2 guard rings, which has to be compared with the dozen of guard rings for non active edge sensors like the ATLAS pixel sensors~\cite{atlasid2}. The dead area at the border of the sensor (the distance between the  last pixel and the border of the sensor) is then considerably reduced, from about 1~mm for traditional sensors to less than 100 $\mu m$  for active edge sensors.
Another feature of this production is the use of a temporary metal line 
which shorts all the pixel together, and  bias them all during the electrical tests, before bump-bonding. The traditional way to do this is to design a permanent structure on the sensors, namely a punch-through grid, which led to some inefficiencies in the charge collection after bump bonding.
The performances of the active edge technology and of the temporary metal line were studied during testbeam and a  detailed presentation of this production can be found in \cite{active_article}.
\subsection{Second production: thin sensors}

The second production~\cite{dallabetta} consists of 100 $\mu m $ and 130 $\mu m $ thick sensors. Such thin sensors are more suitable in hard radiation environment as they are less sensitive to charge trapping.
As the first sensors production, the pixel pitch is $250\mu m \times 50 \mu m$.
This production is also the first production of FBK and LPNHE on 6-inches wafers, the previous one was produced on 4-inches wafers.
The number of guard rings is 2 or 5.
Sensors were also treated against electrical discharge using a BCB passivation coating.
Some of these sensors were irradiated up to a peak fluence of 1 $\times 10^{16} n_{eq}/cm^2$ at CERN IRRAD facility with a beam of 24 GeV/c momentum protons. The irradiation beam profile was a gaussian with a peak at 1 $\times 10^{16} n_{eq}/cm^2$ and a FWHM of 20mm$\times$ 20mm. Those sensors have then been tested on beam, both at DESY and at CERN-SPS. 
\subsection{Third production: thin and active edge sensors}

\begin{figure}[htbp]
\centering

\includegraphics[scale=0.34]{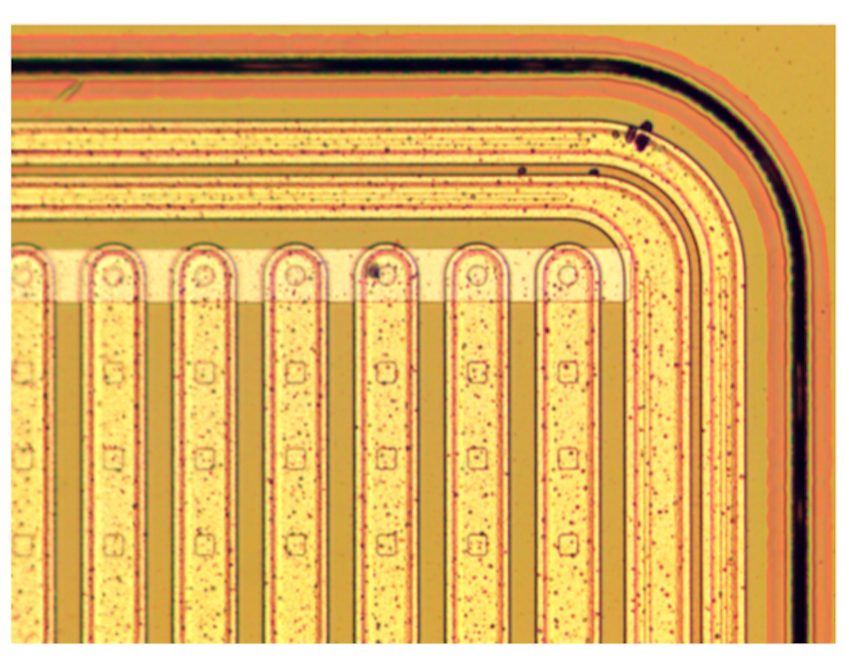}
\includegraphics[scale=0.34]{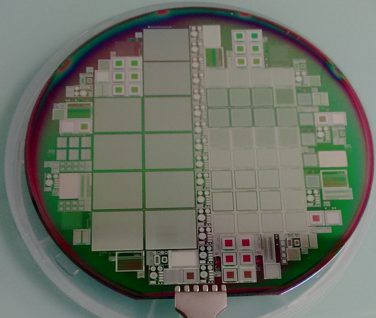}
\includegraphics[scale=0.7]{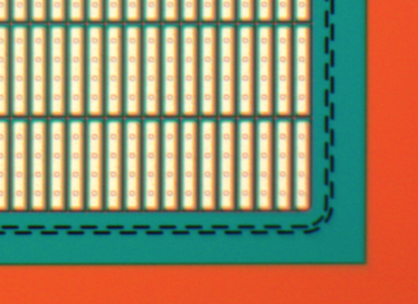}
\caption{\label{fig:2} Details of the three productions. On the left, one can find a photo of a corner of an active edge sensor with 2 guard rings from the first production; in the middle is shown a photo of the 6 inches wafer from the second production of thin sensors; the figure on the right is a zoom on the staggered edge design of the third production.}
\end{figure}

This last  production~\cite{ronchin} combines the two technologies previously used.\footnote{ Some wafers have been sent to IZM to bump bond sensors to FEI4b read-out chips.}
Two bulk thicknesses are considered: 100 $\mu m $ and 130 $\mu m $ as the silicon substrates were the same as the second production. The distance between the trench and the last pixel has been reduced down to 50 $\mu m$, while the number of guard rings is at maximum 1.
The design of the active edge is different than the one used in the first production, it consists of a segmented trench as shown in figure \ref{fig:2}.

\section{Testbeam results}

The first two productions were tested on beam both at DESY with 4 GeV electrons and at CERN SPS with 120 GeV pions. Sensors of the second production were irradiated in two steps at CERN. Data are available for them unirradiated, irradiated at $3\times 10^{15} n_{eq}/cm^2$ and at $1 \times 10^{16} n_{eq}/cm^2$. A copy of the Eudet telescope~\cite{eudet} was used for track reconstruction.
The track reconstruction consists of a set of algorithms, implemented in the EUTelescope framework~\cite{eutelescope}, to process raw data into tracks.
At the end of the process a ROOT~\cite{root} file is created containing basic observables  ready to be analyzed in the data analysis framework, TBmon2 software~\cite{tbmon2}. 
TBmon2 allows studying the quantities discussed below. 
\subsection{Quantities of interest}

\subsubsection*{Configuration of sensors}
Prior to the data taking, sensors have to be carefully configured in terms of threshold and gain. 
When choosing the threshold, a compromise has to be found between a high threshold, which decreases the number of noise hits but decreases the signal efficiency as well, and a low threshold, with opposite effects. After collection by the electrode, the signal is digitized and quantified into Time over Threshold (ToT) by the front-end chip FEI4b~\cite{fei4}. 
 During the tuning of the electronics, the correspondence between ToT value and input charge is calibrated.

 The threshold target depends on the thickness of the sensor and on the fluence it has been exposed to. For an unirradiated 200 $\mu m$ thick sensor of the first production, a typical threshold is 1400~e, which corresponds to a tenth of the expected most probable value signal amplitude due to a minimum ionizing particle (MIP) crossing the sensor at normal incident angle. 
For an irradiated sensor of the second production, due to charge trapping effect and its reduced thickness, the charge collected is lower and so the threshold is between 700e and 1000e; going under 700e can be problematic because of the potentially large noise of irradiated sensors.

\subsubsection*{Global hit efficiency}

The global hit efficiency is defined as the fraction of reconstructed tracks crossing a sensor that have an associated hit in that sensor. The hit efficiency is studied as a function of threshold and bias voltage. 

\subsubsection*{ Charge collection efficiency}
The cluster ToT distribution can be fitted with a Landau convoluted with a gaussian shape. It is interesting to investigate the most probable value (MPV) of this distribution as a function of the bias voltage and the fluence. 
Due to the charge trapping effect, the charge collection efficiency drops sharply with the fluence. The charge collection efficiency increases with the bias voltage, so reaching the maximum MPV gives indication on the maximum signal amplitude that a sensor can reach after a certain fluence.

\subsection{Active edge sensors performances}

The FBK-LPNHE edgeless sensors have demonstrated excellent performance in terms of hit efficiency, especially in the edge area, reaching an efficiency of $97\%$ up to 70 $\mu m$ from last pixel. The use of temporary metal to bias the sensor before bump-bonding has also shown its usefulness, resulting in a homogeneous in-pixel hit efficiency. All the results from this production are discussed in detail in a dedicated article~\cite{active_article} .

\subsection{Thin sensors performances}
A 130 $\mu m$ thick sensor of the second production has been tested at three different fluences: unirradiated, irradiated at $3\times 10^{15} n_{eq}/cm^2$ and at $1 \times 10^{16} n_{eq}/cm^2$.
One of the goal of those irradiated thin sensors is to prove that at least a 97$\%$ hit efficiency in harsh radiation environment is retained. To fulfill this requirement, the sensor has to be capable to reach a high bias voltage without triggering any discharge, hence the sensor is coated with a thin passivation layer (BCB). In the following, the charge collection efficiency and hit efficiency of such sensors are investigated. 
\subsubsection{Charge collection efficiency}
\begin{figure}[htbp]
    \centering
\includegraphics[scale=0.4]{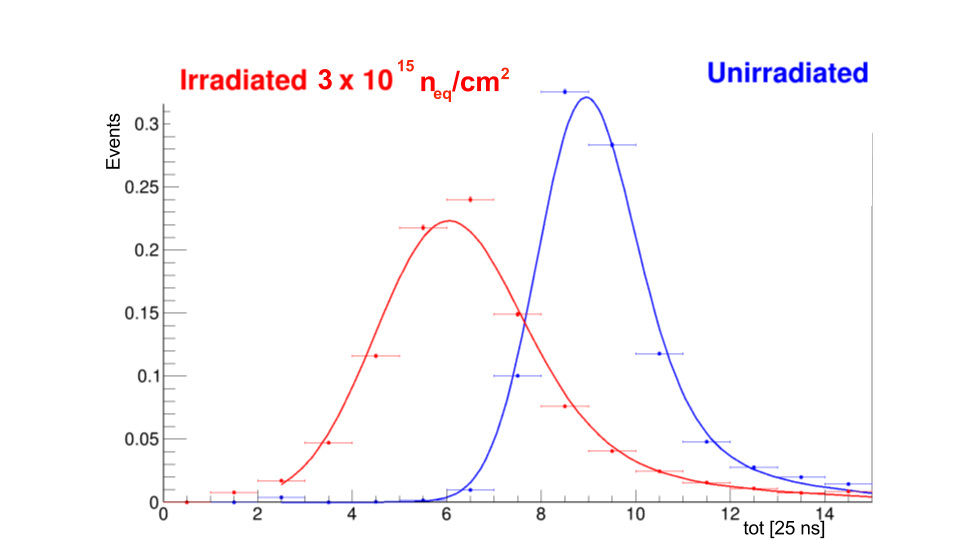} 
\caption{\label{fig:5} ToT distribution for thin unirradiated sensor (blue) biased at 150~V and for thin irradiated at $3\times 10^{15} n_{eq}/cm^2$ sensors biased at 600~V (red)  }
\end{figure}

As shown on figure \ref{fig:5} which presents the ToT distribution for an unirradiated sensor and an irradiated sensor at $3\times 10^{15} n_{eq}/cm^2$, the irradiation of the sensor shifts the ToT distribution towards lower values as some charges are trapped in the bulk due to new states introduced in the band gap by radiation damage.\\ The configuration for both sensors was: threshold at 1000~electrons and 6~ToT corresponds to 6000~electrons; the unirradiated sensor was biased at 150V and the irradiated one at 600~V.\\
The  irradiation at $3\times 10^{15} n_{eq}/cm^2$ of 130 ~$\mu m$ thick sensor reduces the signal by 3 ToT units, the MPV before irradiation is about 9~ToT and after is close to 6~ToT, so a reduction of about 30$\%$ in the charge collection efficiency is observed. As the charge to ToT conversion is not linear, it is non trivial to extract the charge value of the MPV for the unirradiated sensor. For the sensor irradiated at $3\times 10^{15} n_{eq}/cm^2$, the MPV of the ToT distribution is really close to the calibration value (6000~electrons @~6~ToT) so the amount of induced charge is of the order of 6000 electrons.

\begin{figure}[htbp]
    \centering
\includegraphics[scale=0.4]{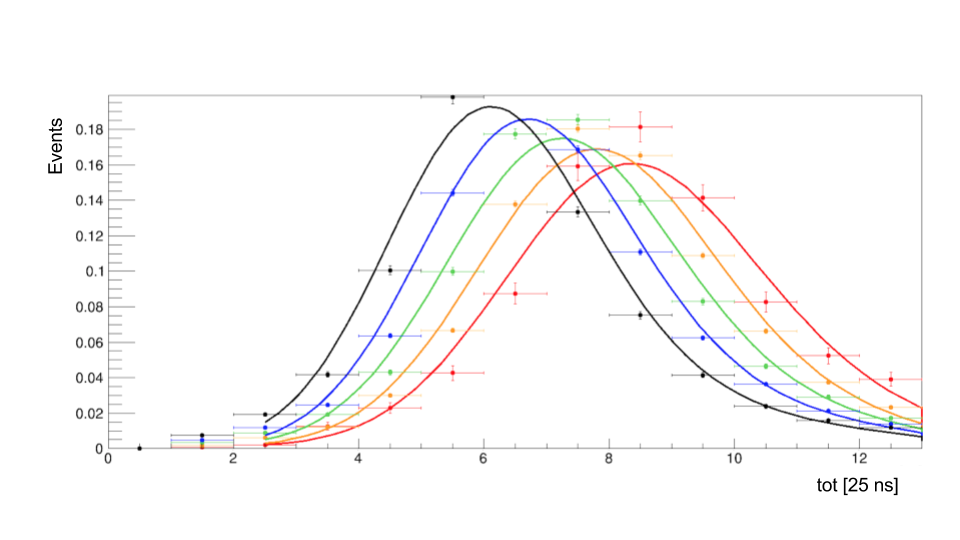} 
\caption{\label{fig:6} ToT distribution for thin sensor irradiated at $1 \times 10^{16} n_{eq}/cm^2$.5 bias voltages between 400~V and 600V were considered (black 400~V, blue 450~V, green 500~V, orange 550~V, red 600~V). All the distribution are fitted by a gaussian convoluted with a landau and the MPV of the ToT distribution is extracted  }
\end{figure}

The figure \ref{fig:6} shows the ToT distribution of the thin sensor irradiated at $1 \times 10^{16} n_{eq}/cm^2$ for 5 different bias voltages, from 400~V to 600~V with a step of 50~V.
The threshold was 700~electrons and the ToT to charge calibration was 8 ToT corresponding to 4000~electrons.
All the distributions were fitted with a Landau convoluted with a gaussian allowing the determination of the most probable value.  At 400~V the MPV is 73$\%$ of the one at 600~V. 

At 600~V the MPV is $\simeq 8.5$ ToT: taking into account the calibration, it means that the induced charge is a bit more than 4000 electrons.
Hence the induced charge at the same bias voltage (600V) has dropped from 6000 electrons to a bit more than 4000 electrons, going from $3\times 10^{15} n_{eq}/cm^2$ to $1 \times 10^{16} n_{eq}/cm^2$. By comparison with unirradiated sensor, this means that the collection efficiency is roughly divided by 2 after an irradiation of $1 \times 10^{16} n_{eq}/cm^2$.
\subsubsection{Global hit efficiency}
\begin{figure}[htbp]
\centering 
\includegraphics[scale=0.4]{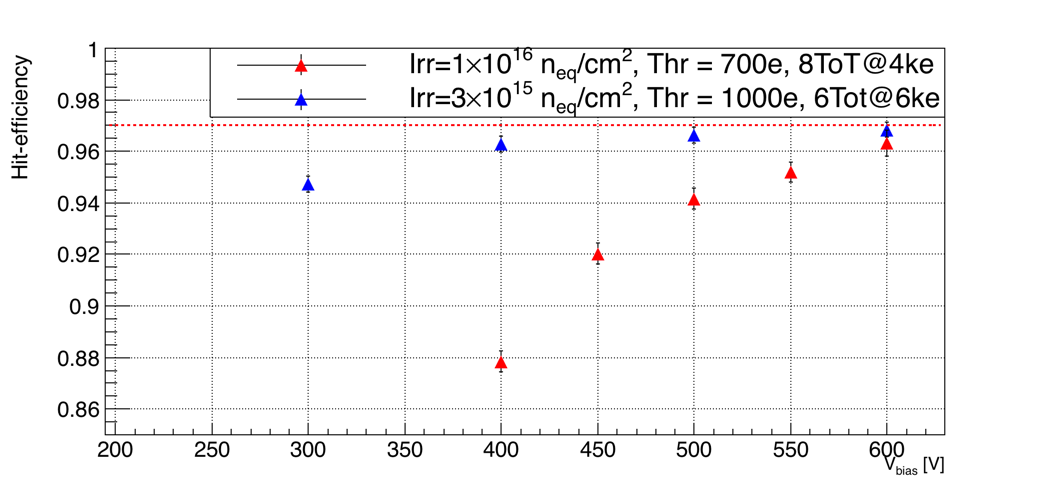} 
\caption{\label{fig:7} Hit efficiency for thin irradiated sensors. The red triangles are for sensor irradiated at $1\times 10^{16} n_{eq}/cm^2$ and the blue ones at $3\times 10^{15} n_{eq}/cm^2$. Threshold and gain are indicated in the upper box. }
\end{figure}

 As shown in the figure \ref{fig:7}, the efficiency of the thin sensors is really close to the 97$\%$ ATLAS requirement. At $3\times 10^{15} n_{eq}/cm^2$ and 600~V, the efficiency reaches  97$\%$. At $1\times 10^{16} n_{eq}/cm^2$ and 600~V, the efficiency is $96.32 \pm 0.5\%$, quite close to the 97$\%$ ATLAS requirement. To reach those good efficiency results it is mandatory to choose a low threshold, because the charge collection efficiency is really degraded by the fluence, as shown in the previous part. The results in terms of efficiency indicate that those sensors are serious candidates for the outer part of the tracker. 
\newpage
\section{Conclusions and outlook}

The two productions of planar pixel sensors tested on beam have both shown great performance in terms of efficiency.\\
The active edge process of FBK has proven its efficiency, the edge region is efficient above $97\%$ up to 70 $\mu m$ from last pixel, active edge sensors are then capable of granting an increase of the geometrical acceptance of the ATLAS modules.
The temporary metal technology allows the sensors to reach a very homogeneous efficiency, hence it is considered an excellent biasing option for the new tracker sensors.\\
The thin sensor production reached the 97$\%$ hit-efficiency requirement when exposed at a fluence of $3\times 10^{15} n_{eq}/cm^2$ and the performance at $1\times 10^{16} n_{eq}/cm^2$ are really close to this requirement. The charge collection efficiency has also been investigated and a drop of approximately 50$\%$ between the unirradiated and the irradiated at $1\times 10^{16} n_{eq}/cm^2$ conditions have been observed.\\
The third production, which combines the active edge technology on thin sensor production is being bump-bonded and will be ready to be put on the beam by the beginning of 2018.

\acknowledgments

Productions 2 and 3 are supported by the Italian National Institute for Nuclear Research (INFN), Projects ATLAS, CMS, RD-FASE2 (CSN1) and by AIDA-2020 Project EU-INFRA Proposal. 
The authors want to thanks the CERN irrad team for helping with the irradiation of the sensors.



\begin{thebibliography}{99}
\bibitem{atlas}
ATLAS, \emph{The ATLAS Experiment at the CERN Large Hadron Collider}, JINST 3 (2008) S08003.
\bibitem{atlasid1}
ATLAS Collaboration, \emph{The ATLAS Inner Detector commissioning and calibration},
Eur. Phys. J. C70 (2010) 787
\bibitem{atlasid2}
G. Aad et al., \emph{ATLAS pixel detector electronics and sensors}, JINST 3 (2008) P07007


\bibitem{ibl}
M Capeans et al., \emph{ATLAS Insertable B-Layer Technical Design Report}, tech. rep. CERN-LHCC-2010-013. ATLAS-TDR-19, 2010, https://cds.cern.ch/record/1291633.

\bibitem{hllhc}
The HL-LHC project, http://hilumilhc.web.cern.ch/about/hl-lhc-project

\bibitem{rd53}
RD53 collaboration, https://rd53.web.cern.ch/RD53/

\bibitem{strips} ATLAS collaboration, Technical Design Report for the ATLAS Inner Tracker Strip Detector, CERN-LHCC-2017-005 ATLAS-TDR-025, Apr 2017, https://atlas.web.cern.ch/Atlas/GROUPS/PHYSICS/UPGRADE/CERN-LHCC-2017-005/

\bibitem{bomben}

M.~Bomben et al.,\emph{Development of Edgeless n-on-p Planar Pixel Sensors for future ATLAS Upgrades},
Nucl. Instr.  and Meth.A, 712, 2013, 41-47,10.1016/j.nima.2013.02.010

\bibitem{fei4}
M.~Garcia-Sciveres et al, \emph{ The {FE-I4} pixel readout integrated circuit}, Nucl. Instr. and Meth. A,636,S155-S159,2011,

\bibitem{active}
M.Povoli et al, \emph{Development of planar detectors with active edge},
Nucl. Instr. and Meth. A,658 (2011)






\bibitem{active_article}
 M.Bomben, A.Ducourthial et al, \emph{Performance of active edge pixel sensors},  JINST 12 P05006 (2017)
 
\bibitem{dallabetta}
 G.-F. Dalla Betta  et al, \emph{The INFN-FBK Phase-2 R$\&$D program}, Nucl. Instr. and Meth. A,824,388 - 391,2016
\bibitem{ronchin} 
S. Ronchin, et al, \emph{Edgeless planar pixel sensors with ATLAS and CMS designs produced by FBK-CMM},12th Trento Workshop on Advanced Silicon Radiation Detectors, 2017

 
\bibitem{eudet}
H.Jansen, \emph{Performance of the EUDET-type beam telescopes}, DESY-16-055,
arXiv:1603.09669v2 [physics.ins-det] 10 May 2016
\bibitem{eutelescope}
Eutelescope, http://eutelescope.web.cern.ch/ .
\bibitem{root}
ROOT Data Analysis Framework, https://root.cern.ch,
\bibitem{tbmon2}
Tbmon2, https://bitbucket.org/TBmon2/tbmon2/overview.











\end{thebibliography}
\end{document}